\documentclass[10pt,preprint]{aastex}
\begin{document}

\title{DISCOVERY OF A CANDIDATE PROTOPLANETARY DISK AROUND THE EMBEDDED
SOURCE IRc9 IN ORION\altaffilmark{1}}

\author{Nathan Smith\altaffilmark{2} and John Bally}

\affil{Center for Astrophysics and Space Astronomy, University of
Colorado, 389 UCB, Boulder, CO 80309; nathans@casa.colorado.edu}

\altaffiltext{1}{Based on observations obtained at the Gemini
Observatory, which is operated by the Association of Universities for
Research in Astronomy, Inc., under a cooperative agreement with the
NSF on behalf of the Gemini partnership: the National Science
Foundation (US), the Particle Physics and Astronomy Research Council
(UK), the National Research Council (Canada), CONICYT (Chile), the
Australian Research Council (Australia), CNPq (Brazil), and CONICET
(Argentina).}

\altaffiltext{2}{Hubble Fellow}

\begin{abstract}

We report the detection of spatially-extended mid-infrared emission
around the luminous embedded star IRc9 in OMC-1, as seen in 8.8, 11.7,
and 18.3 $\micron$ images obtained with T-ReCS on Gemini South.  The
extended emission is asymmetric, and the morphology is reminiscent of
warm dust disks around other young stars.  The putative disk has a
radius of roughly 1$\farcs$5 (700 AU), and a likely dust mass of
almost 10 Earth masses.  The infrared spectral energy distribution of
IRc9 indicates a total luminosity of $\sim$100 L$_{\odot}$, implying
that it shall become an early A-type star when it reaches the main
sequence.  Thus, the candidate disk around IRc9 may be a young analog
of the planetary debris disks around Vega-like stars and the disks of
Herbig Ae stars, and may provide a laboratory in which to study the
earliest phases of planet formation.  A disk around IRc9 may also add
weight to the hypothesis that an enhanced T~Tauri-like wind from this
star has influenced the molecular outflow from the OMC-1 core.

\end{abstract}


\section{INTRODUCTION}

One of the most spectacular bipolar protostellar outflows known is the
system of H$_2$ ``fingers'' emanating from the BN/KL region in the
OMC-1 cloud core (Allen \& Burton 1993; Schild et al.\ 1997; Kaifu et
al.\ 2000), at a distance of roughly 460 pc (Bally et al.\ 2000).
Among the embedded infrared (IR) sources associated with the BN/KL
complex, the bright source IRc9 (Downes et al.\ 1981; Wynn-Williams \&
Becklin 1974) is relatively isolated, located $\sim$30$\arcsec$ north
of the rest of BN/KL (see Fig.\ 1$a$).  The H$_2$ outflow from BN/KL
appears highly asymmetric, with the northwest/blueshifted part of the
flow being brighter and more extended.  This asymmetry is not easily
explained by the tilt angle and corresponding potential extinction of
the redshifted flow, since observations of H$_2$ at both 2.12
$\micron$ and 12.3 $\micron$ show the same degree of asymmetry (Beck
1984).  Suspiciously, IRc9 is found near the H$_2$ emission peak
within the brighter northwest part of the bipolar outflow.  This may
be a simple coincidence, but Beck (1984) has proposed the alternative
view that IRc9 may have had a direct role in influencing the
asymmetric appearance the BN/KL outflow.  Specifically, Beck (1984)
suggests that a T~Tauri-like wind from IRc9, if it exists, may have
helped create a lower density region, allowing the northwest part of
the BN/KL outflow to escape to larger distances with higher speeds
than the southeast counterflow.  A direct link between IRc9 and the
H$_2$ outflow is supported by unusual kinematics seen in spectra of
the 12.28~$\micron$ H$_2$ emission line, which reveal that the most
complex line shapes are found in the region near IRc9 (Beck et al.\
1982).  Thus, while IRc9 does not give rise to the H$_2$ outflow
itself, it is a good candidate for influencing the outflow's
asymmetry.

In this Letter we report the detection of a spatially-extended
asymmetric dust envelope around IRc9, probably representing a large
circumstellar disk.  The presence of a disk would add weight to the
hypothesis that IRc9 is a young star with an active outflowing wind
analogous to T~Tauri stars, as required in the scenario proposed by
Beck (1984).  However, even if Beck's interpretation is incorrect, the
extended emission we report here suggests that IRc9 is an interesting
source in its own right because it may be surrounded by a very young
protoplanetary disk.  The integrated luminosity of IRc9 suggests that
when it eventually reaches the main-sequence, it will be an early
A-type star.  Therefore, the disk around IRc9 may also provide an
example of the youngest phases in a planet-forming disk that will
eventually evolve into a debris disk analogous to those seen around
Vega-like stars.

\section{OBSERVATIONS}

On 2004 Jan 25, 26, and 27 we used T-ReCS\footnote{{\url
http://www.gemini.edu/sciops/instruments/miri/MiriIndex.html}.} on
Gemini South to obtain 8.8, 11.7, and 18.3 $\micron$ images of IRc9,
located about 30$\arcsec$ north of the BN/KL nebula in Orion (see
Fig.\ 1$a$). T-ReCS is Gemini South's facility mid-IR imager and
spectrograph with a 320$\times$240 pixel Si:As IBC array, a pixel
scale on the 8m telescope of 0$\farcs$089, and a resulting
field-of-view of 28$\farcs$5$\times$21$\farcs$4.  The observations
were taken with a 15$\arcsec$ east-west chop throw, and a chopping
frequencey of a few Hz (different for each filter).  In each filter, 4
spatially-offset groups of chop-nod pairs (each composed of 10
individual pairs) were combined to make the final image, using the
central peak of IRc9 itself for spatial alignment.  These observations
were part of a larger wide-field mosaic of Orion, and additional
T-ReCS images of Orion from this same dataset were presented in an
earlier paper (Smith et al.\ 2004).

Figures 1$b$, $c$, and $d$ show the resulting 8.8, 11.7, and
18.3~$\micron$ images of IRc9.  The images were flux calibrated using
observations of the secondary standard star HD~32887, adopting the
values tabulated by Cohen et al.\ (1999).  Absolute photometric
uncertainty is probably dominated by the $\sim$5\% uncertainty in the
calibration, rather than the 1$\sigma$ noise-equivalent flux density
in the background for an aperture of the same size given in Table 1.
The nights were photometric during the 8.8 and 11.7 $\micron$
observations, and the sky was photometric for the early part of the
18.3 $\micron$ mosaic when IRc9 was observed, although conditions
deteriorated later in the night.  Background-subtracted photometry for
IRc9 is listed in Table 1, measured in a 5$\farcs$3 diameter aperture
to include all the extended structure visible in images.

To clarify the extended structure around IRc9, we subtracted a scaled
point spread function (PSF) from the images (Figs.\ 1$e$, $f$, and
$g$), and Table 1 also lists the corresponding flux of the star that
was subtracted and the residual emission left after the PSF
subtraction.  The PSF used for subtraction in each filter was the
corresponding observation of the standard star, obtained $\sim$10 min
before the 8.8 and 18.3 $\micron$ observations, and $\sim$25 min prior
to the 11.7 $\micron$ observation of IRc9.  The empirical PSF was
scaled and subtracted from each of the IRc9 images in a somewhat
subjective manner -- chosen to minimize residual emission near the
star without severely over-subtracting the central PSF (i.e. to avoid
creating negative flux values).  Uncertainty in the residual flux may
be as high as 30\% at 8.8 $\micron$ where the central star was bright
compared to the total extended flux, but the photometric uncertainty
in the residual flux at 18.3 $\micron$ is comparable to the $\sim$5\%
calibration uncertainty because the extended emission constitues the
majority of the flux from IRc9 at that wavelength.  Obviously the
uncertainty in the residual structure is highest near the position of
the star where small differences in the PSF may leave the most severe
artifacts (especially at the shorter wavelengths).  However, structure
beyond 1\arcsec\ from the star's position is much more reliable, and
the uncertainty there is comparable to the statistical uncertainty
(the lowest contour is drawn at 3$\sigma$ above the background in
Figs.\ 1 $efg$).

\section{RESULTS}

Figure 1 shows that IRc9 is an extended object, with a size of
3--4$\arcsec$, elongated roughly along P.A.$\simeq$120$\arcdeg$.  This
elongated structure is evident both before and after subtraction of
the central point source.  Given the shape of the extended emission,
it is likely that the central star powering IRc9 is surrounded by a
large dusty circumstellar disk, with the polar axis oriented at
P.A.$\simeq$30$\arcdeg$.  The residual emission after the PSF
subtraction is spatially-resolved along both the minor and major axes
of the nebula; if this is a tilted circular disk then the difference
between the major and minor dimensions suggests that its polar
axis is inclined from the plane of the sky by perhaps 30$\arcdeg$.

Is this emission really from a disk?  An alternative view might be
that this represents externally-illuminated material in the OMC-1
cloud core, as is the case for many of the complex dust structures
seen in the BN/KL nebula (e.g., Shuping et al.\ 2004).  While this
interpretation is difficult to refute conclusively with the present
dataset, we consider it less likely than the disk hypothesis because
IRc9 is relatively isolated from the BN/KL core (see Fig.\ 1a), and no
other point source out of 83 that we detected in our larger mosaic of
the inner Orion nebula shows similar extended structure.  The diameter
of this putative disk (1400--1800 AU) is comparable to or somewhat
larger than the largest proplyd silhouette disks in Orion (Bally et
al.\ 2000; Shuping et al.\ 2003; Smith et al.\ 2004a), and is similar
to the sizes of several disk candidates seen in the Carina nebula
(Smith et al.\ 2003).

IRc9's disk is brighter on the southeast side, with the brightness
asymmetry being most pronounced (roughly a factor of 2) at the longest
wavelength.  Interestingly, the bright part of the disk faces toward
the heart of the BN/KL complex.  If this apparent asymmetry arises
because a portion of IRc9's disk is heated externally by the BN/KL
central engine or by interaction with its molecular outflow, this
would provide evidence that the two objects are indeed at the same
distance.

The spectral energy distribution (SED) of IRc9 is shown in Figure 2,
including the total flux, the subtracted PSF, and the residual
emission from Table 1.  This SED obviously cannot be reproduced with a
single-temperature dust model, so Figure 2 shows a simple fit using
three individual temperature components at 750, 170, and 115~K.  These
individual components are guided by the spatially-resolved photometry
in Table 1, plotted in Figure 2.  The 750 and 170 K components
together constitute the central PSF source that was subtracted, while
the cooler 115~K dust component matches the flux from the residual
emission after the PSF was subtracted (Figs.\ 1$e$, $f$, and $g$).
The central point-like source is probably a mix of reddened
photospheric emission and a range of hot dust temperatures in a
compact ($R<$160 AU) disk around the star.  The 115~K component
represents optically-thin emission from cooler dust in the extended
circumstellar disk.  The integrated luminosity of each component and
the total of all three are listed in Table 2.  If the dust grains are
small ($a<$0.2~$\micron$) the dust mass required to emit this IR
luminosity can be expressed independent of the grain radius and
emissivity (see Smith \& Gehrz 2005), so that $M_{\rm
dust}=[(100\rho)/(3\sigma T^6)]\,L_{\rm IR}$.  This relation was used
to derive the dust masses for each component in Table 2.  If the
dominant dust grains are large ($a>$1~$\micron$) then the mass
estimate is more uncertain and far-IR or submillimeter measurements
are probably needed.  (However, we do not see an obvious reason why
large grains would dominate the grain size distribution in the outer
disk for this embedded source, which is shielded from the strong
external UV radiation field of the Trapezium.)

Most of the mass resides in the coolest extended component (about 8
$M_{\earth}$), indicating that the disk around IRc9 provides a
significant reservoir out of which planets may potentially form. Note,
however, that since we ignored extinction and since optically-thick
regions may be present in the inner disk, the mass estimates for the
750 and 170 K components are really a lower limit to the mass of the
warm inner parts of the disk.  The dust temperature of 115~K in the
extended disk is consistent with the equilibrium temperature for small
($a\simeq$0.1~$\micron$) grains at a separation of 1$\arcsec$ (460 AU)
from a central source with a luminosity of $\sim$100--200 L$_{\odot}$
(assuming $Q_{\rm abs}/Q_{\rm em}\approx$100 for small grains; see
Smith \& Gehrz 2005 and references therein).  This is roughly
consistent with or somewhat higher than the integrated IR luminosity
of IRc9 (Table 2), perhaps suggesting that some of the source
luminosity may escape in directions not intercepted by the disk.

\section{DISCUSSION}

If the extended mid-IR emission around IRc9 really does represent a
circumstellar disk, then it has two important ramifications.  An
active accretion disk around IRc9 would suggest that it is a very
young pre--main-sequence star.  This would add plausibility to the
hypothesis that it has a strong T~Tauri-like mass-loss wind that could
have potentially affected the appearance of the bipolar molecular
outflow from BN/KL (Beck 1984), although this hypothesis remains
speculative.  Additional evidence that IRc9 may be a very young object
comes from its Br$\gamma$ emission (Beck 1984).  If IRc9 is embedded
in the outflow from the BN/KL region, could it survive?  Embedded in
the OMC-1 cloud, IRc9 is shielded from photoablation by the intense UV
radiation of the Trapezium, where proplyd disks are thought to survive
for 10$^5$ yr or more (O'Dell 1998), whereas the dynamical time of the
BN/KL outflow is only about 1000 yr (Lee \& Burton 2000; Doi et al.\
2002).  Estimating the disk's ability to survive dynamical interaction
with shocks caused by the BN/KL outflow itself is more difficult, and
probably requires a numerical investigation beyond the scope of this
Letter.

Even more tantilizing is that a young disk around IRc9 may provide a
laboratory to study the earliest phases of planet formation.  Recent
studies of several A-type stars have revealed extended thermal-IR
emission from dust in planetary debris disks.  Some well-known
examples are the disks around $\beta$~Pic (A5 V; Wahhaj et al.\ 2003;
Weinberger et al.\ 2003), HR~4796A (A0 V; Koerner et al.\ 1998;
Jayawardhana et al.\ 1998; Schneider et al.\ 1999), Fomalhaut (A3 V;
Holland et al.\ 2003), HD~141569 (B9.5 Ve; Weinberger et al.\ 1999),
and Vega itself (Holland et al.\ 1998).  These Vega-like stars are
typically a few to tens of Myr old -- older than expected timescales
for planet formation -- so mature planets have presumably already
formed, and their disks are probably the remnants of the planet
formation process.  The dust in these disks is thought to be
replenished from collisions of planetessimals (Backman \& Paresce
1993).  These debris disks typically have radii somewhat smaller than
that of IRc9, but significantly lower masses.  It is intriguing that
the apparent size, mass, and the mildly asymmetric structure of the
material around IRc9 are reminiscent of the disk and envelope around
the Herbig Ae star AB Aurigae (Fukugawa et al.\ 2004).

The integrated luminosity of IRc9 is roughly 100 L$_{\odot}$, which
places it on a pre--main-sequence track for a star that will
eventually have M$_{\rm ZAMS}\simeq$3--4 M$_{\odot}$ and an early A
spectral type (e.g., Marconi \& Palla 2004).  Thus, the disk around
IRc9 may be an early analog of these more evolved debris disks, and so
we speculate that it is a good candidate for a very young
protoplanetary disk around a Vega-like star.  If true, IRc9 may be a
valuable laboratory for studying the earliest phases of planet
formation.  In any case, IRc9 certainly deserves closer scrutiny.

\acknowledgments  \scriptsize

Support for N.S.\ was provided by NASA through grant HF-01166.01A from
the Space Telescope Science Institute, which is operated by the
Association of Universities for Research in Astronomy, Inc., under
NASA contract NAS5-26555.  We thank an anonymous referee for several
suggestions that improved the presentation of our results.


\begin{deluxetable}{llccc}
\tabletypesize{\scriptsize}
\tablecaption{Observations and Photometry of IRc9}
\tablewidth{0pt}
\tablehead{
\colhead{Parameter} &\colhead{Units} 
  &\colhead{8.8 $\micron$} &\colhead{11.7 $\micron$} &\colhead{18.3 $\micron$} }
\startdata
filter $\Delta\lambda$	&$\micron$	&0.78	&1.13	&1.5	\\
exp.\ time		&sec.\		&72	&72	&72	\\
1$\sigma$ F$_{\nu}$	&mJy		&8.3	&10.5	&66	\\
F$_{\nu}$ total		&Jy		&5.85	&9.46	&30.5	\\
F$_{\nu}$ star		&Jy		&5.25	&6.85	&10.9	\\
F$_{\nu}$ PSFsub	&Jy		&0.60	&2.61	&19.6	\\
FWZI\tablenotemark{a}	&arcsec		&3.0	&3.6	&4.0	\\
\enddata
\tablenotetext{a}{Size of the major-axis diameter of the
PSF-subtracted emission at the lowest contour level.}
\end{deluxetable}

\begin{deluxetable}{lcc}
\tabletypesize{\scriptsize}
\tablecaption{Luminosity and Dust Mass}
\tablewidth{0pt}
\tablehead{
\colhead{Component} &\colhead{L$_{\rm IR}$}  &\colhead{M$_{\rm dust}$} \\
\colhead{\ }        &\colhead{(L$_{\odot}$)} &\colhead{(M$_{\earth}$)} }
\startdata
750 K		&40	&8.8$\times$10$^{-5}$	\\
170 K		&14	&0.22	\\
115 K		&48	&8.1	\\
Total		&102	&8.3	\\
\enddata
\end{deluxetable}

\begin{figure}
\epsscale{0.95}
\plotone{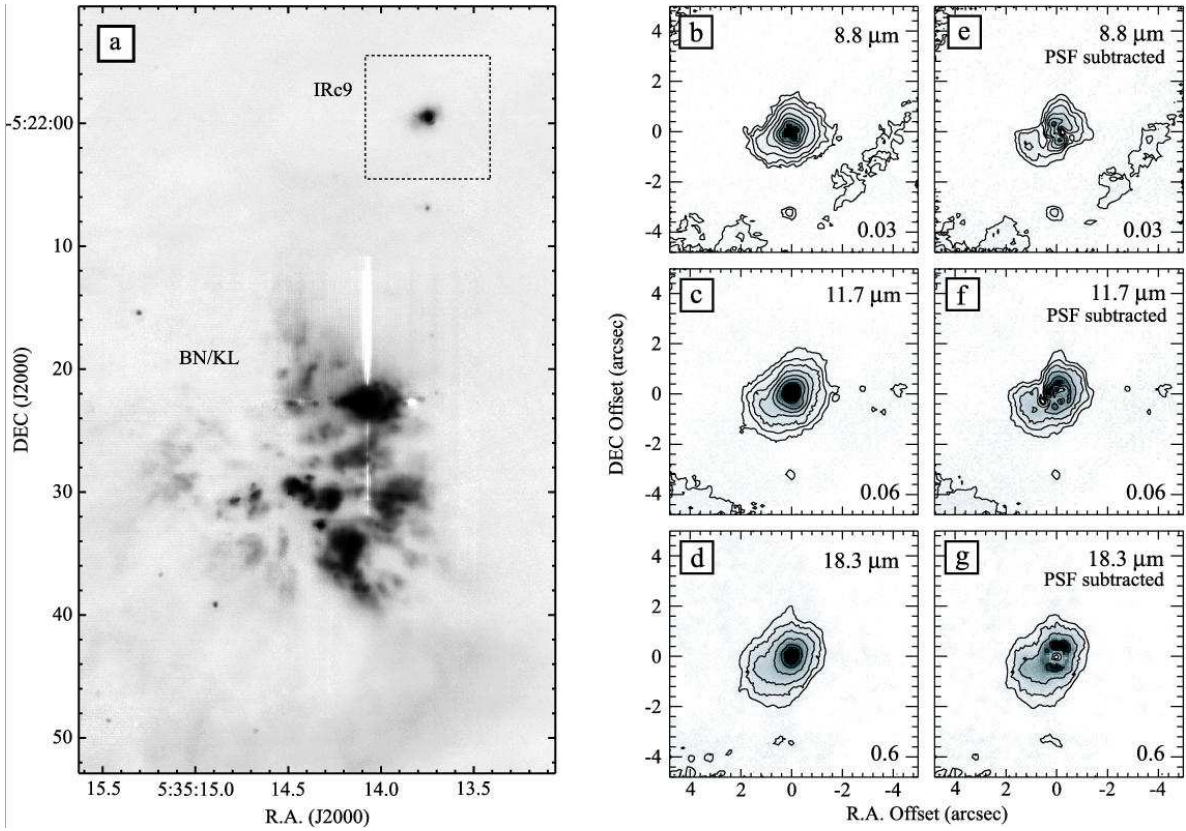}
\caption{(a) Northern portion of our 11.7 $\micron$ image mosaic of
the Orion nebula. (b, c, d) Raw images of IRc9 at 8.8, 11.7, and 18.3
$\micron$, respectively, corresponding to the box in Panel $a$.  (e,
f, g) Same as neighboring images, except that an observed PSF has been
subtracted.  In Panels $b$--$g$, the lowest contour level in Jy
arcsec$^{-2}$ is given in the lower right corner, and each subsequent
contour level is a factor of 2 above the previous one.}
\end{figure}

\begin{figure}
\epsscale{0.5}
\plotone{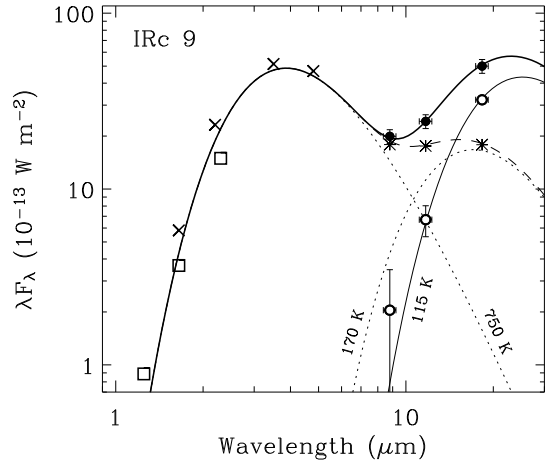}
\caption{The spectral energy distribution of IRc9.  Filled circles are
the total flux in a 5$\farcs$3 diameter aperture measured in T-ReCS
images, unfilled circles are the PSF-subtracted flux, and stars
represent the PSF flux that was subtracted.  Unfilled squares are
2MASS data, and X's are photometric measurements by Wynn-Williams \&
Becklin (1974).  The various curves are individual Planck functions
fit to the data with a $\lambda^{-1}$ emissivity, or the sum of those
Planck functions.  The thin solid line represents the extended 115~K
emission, the dashed line is the central point-like source (the sum of
the dotted curves for the 750 and 170 K components), and the heavy
solid line is the total emission.}
\end{figure}

\end{document}